\documentclass[%
reprint,
 amsmath,amssymb,
 aps,
 prd,
floatfix,
]{revtex4-2}

\usepackage{graphicx}
\usepackage{dcolumn}
\usepackage{bm}
\usepackage{hyperref}

\begin{document}

\preprint{APS/123-QED}

\title{Efficient identification of infected sub-population}

\author{An\v{z}e Slosar}
 \affiliation{Brookhaven National Laboratory, Upton NY 11973}

\date{\today}

\begin{abstract}
  When testing for infections, the standard method is to test each subject individually. If testing methodology is such that samples from multiple subjects can be efficiently combined and tested at once, yielding a positive results if any one subject in the subgroup is positive, then one can often identify the infected sub-population with a considerably lower number of tests compared to the number of test subjects. We present two such methods that allow an increase in testing efficiency (in terms of total number of test performed) by a factor of $\approx$ 10 if population infection rate is $10^{-2}$ and a factor of $\approx$50 when it is $10^{-3}$. Such methods could be useful when testing large fractions of the total population, as will be perhaps required during the current coronavirus pandemic.
\end{abstract}

\maketitle

\section{Introduction}

During the recent coronavirus outbreak in the US, it has been suggested on the Twitter that poor people that are unable to pay for the COVID-19 testing, can instead opt for coughing onto a rich person and wait for them be tested. This has prompted this author to think about how should a set of poor people go about coughing onto a limited number of rich people in order to optimally determine who is infected among them. Somewhat less morbidly, the problem is exact identification of infected people from a pool of $N$ people using fewer than $N$ tests.

Of course, this is only possible if samples from multiple people are somehow combined. In this note we consider a test in which samples from $M$ subjects are combined into a single sample, which tests positive if one more or more constituent subjects are positive. Whether this is viable in practice is beyond the scope of this note, but one would naively expect that it should be possible with testing methodology that relies on detecting trace viral fragments. Since we still need samples from all $N$ subjects, this and related techniques makes sense only if testing rather than collecting samples is the resource limiting step. With these caveats, let us proceed to the calculation.

\begin{table*}[!ht]

\begin{tabular}{|c|c||c|p{3cm}|c|c||c|c|c|c|c||c|}
\hline
$f$ & true infect & Iterations & $M$ & $N_t$ & cost &  $M$ & $K$ & false positives & $N_t$ & cost & theoretical min cost. \\
\hline
$10^{-2}$& 1077 & 8 & 68, 35, 20, 12, 7, 4, 2, 1 & 13461 & 0.13  & 69 & 6 & 1950 & 8700/11727 & 0.09/0.12  & 0.08 \\
$10^{-3}$& 106 & 12 & 692, 357, 196, 125, 77, 48, 30, 19, 12, 7, 3, 1 & 2003 & 0.020  & 693 & 9 & 301 & 1305/1712 & 0.013/0.017  & 0.011 \\
\hline
\end{tabular}
\caption{Results for both methods using a toy example of a perfect test and $f$ of either $10^{-2}$ or $10^{-3}$. $N=10^5$ in both cases. For Divide and Conquer method we show the number of iterations required to converge, the vales $M$ took during these iterations and final cost (number of tests divided by $N$). For Group coding method we show the values of $M$ and $K$ used (given by Eqs.~\ref{eq:1} and \ref{eq:2}, rounded to the nearest integer) and the total number of test used. We give two values of cost: the lower number is without a second pass to weed out false positives. In the last column we give the information theoretical minnimum cost of Eq.~\ref{eq:min}. }
\label{tab:table1}
\end{table*}

\section{Context}
Let $f$ be the overall rate of infections in a population of size $N$. The information content in who is infected and who is not, is given by 
\begin{equation}
    I = N \left( -f\log_2 f - (1-f)\log_2(1-f)\right), \label{eq:min}
\end{equation}
in bits, i.e. it would take in average that many questions with a yes/no answer to uniquely determine who is infected.  Since each testing procedure gives one bit of information, it also sets the theoretical lower bound on the required number of tests.  The numbers are  
$I\sim0.08N$ for $f=10^{-2}$ and $I\sim0.011N$ for $f=10^{-3}$. The lower the population infection rate, the fewer bits of information are needed to describe it.  Of course,  existence of this lower bound does not actually guarantee that a better method exist or is practicable.

By a similar token, an information content from a single test is optimally informative, when there is the same probability of getting a positive or negative answer. Let us combine samples from $M$ subjects and let assume the test if positive is any one of them is positive. The probability of test being negative is $(1-f)^M$ and requiring this to be $1/2$ we get
\begin{equation}
    M = -\left(\log_2 (1-f)\right)^{-1} \label{eq:1}
\end{equation}

So the main trick is to combine subjects in groups of size $M$, so that a test on such group has about the same probability of being positive or negative, which maximizes the information gain from the test. However, we of course need to repeat tests in order to identify the actually infected subjects.
Below we give two example methods.

\subsection{Method 1: Divide and Conquer}

The Method 1 is a simple divide and conquer approach.  We use the estimate of $f$ to make a first pass over the entire population spliting it into $N/M$ groups. As discussed above, approximately half these groups will test negative and the other half are now "concetrated" with effective $f$ approximately double that of full populationa and hence  $M$ approximately half. We repeat the process until $M$ reduces to unity, at which point we individually test the remaining subjects, yielding the infected sub-population.

\subsection{Method 2: Group coding}

Method 2 is somewhat more complicated, but in our simulation tests performs marginally better and has a distinct advantage that it is perfectly parallelizable, at least in the most time-consuming first step, because all groupings are decided in advance.

Again we start by generating groups of size $M$, with a total number of groups give by $N_g = NK/M$, where $K$ is a parameter that controls the number of false positives as discussed below. These groups are such that each subject appears in $K$ groups and no two subjects appear in exactly the same set of groups, i.e. a set of groups uniquely codes a given subject.

We then proceed to test each of these $N_g$ groups. We use these results to assign infected status as follows: each subject is deemed positive if all the outcomes from all of the $K$ groups they belong to are positive and negative otherwise. 

If the subject is actually positive, then all of their groups will test positive, so they will be marked positive. In fact, for a perfect underlying test, there are no false negatives.  On the other hand, if subject is negative, then it will test positive with a probability $\sim(1/2)^K$, because we have arranged $M$ so that each group has equal probability of testing positive or negative. We can make $K$ large enough so that the number of false positives is manageable. Alternatively, we can retest all the positives, which brings the total number of tests to
\begin{equation}
    N_t = N \left( \frac{K}{M} + f + (1-f)\left(\frac{1}{2}\right)^K \right)
\end{equation}
We can now optimize $K$ for smallest $N_t$, giving a surprisingly ugly equation
\begin{equation}
    K = -\log_2\left( \frac{-\log_2(1-f)}{(1-f)\ln 2}\right) \label{eq:2}
\end{equation}

\subsection{Results}

We have coded a toy example of both methods in a Jupyter notebook, which can be found at \url{https://github.com/slosar/infections}. We present results in the Table \ref{tab:table1}.  We see that both methods perform reasonably close to theoretical expectations.  In particular, for $f=10^{-2}$ we can get away with $0.13N$ or $0.13N$ tests compared to the brute force $N$ tests. This is somewhat less efficient that the theoretically optimum number of $0.08N$. For $10^{-3}$, we are again performing worse than the theoretical minimum with $0.020N$ or $0.017N$ test rather than $0.011N$, but still with a very large efficiency gain compared to the brute-force $N$ tests. In particular, if we are willing to live with some false positives (i.e. quarantining a set of unlucky souls) then the number of tests is even lower.

It is likely that methods could be improved further, however at likely diminishing costs.

\section{Conclusions}

In this note we have presented two methods for identifying the infected individuals by using slightly more sophisticated methods than brute-force testing every single sample. 

Both methods are in statistically perfect with no false positives and no false negatives (assuming second pass to weed out false positives in Method 2), but can amplify underlying errors. For example, if method has a false negative rate of $p$, then the false negative rate of Method 2 will be $\sim Kp$ and similar for Method 1. 

Both methods are likely impractical using the current testing methods, because of housekeeping complexity of preparing dividing up and combining samples without making a mess of your laboratory. However, it is conceivable that future testing machines could employ Method 1, by taking $2^N$ samples and directly performing the necessary combinations and divisions internally.

To answer the original question: the poor people should cough onto the insured rich in a way that make the rich person probability of catching infection about 50\%.  Alternatively, countries should adopt medical systems in which no person would need to cough onto anybody. Otherwise, the history will do it for them \cite{10.2307/j.ctv346rs7.23}.

\bibliography{apssamp}

\end{document}